\documentclass[12pt,letterpaper,twocolumn,fleqn]{narms}
\usepackage{subfigure}
\usepackage{epsfig}
\usepackage{timesmt}

\usepackage{amsmath,amssymb,mathtools}
\usepackage{lipsum} % for widetext

\newcommand{\argmin}{\operatornamewithlimits{argmin}}

\usepackage{chicaco}

\begin{document}
\title{Detecting structural breaks in seasonal time series by regularized optimization}
\author{{Bing Wang} \\
{\aff{Department of Mathematics and Computer Science, Clarkson University, Potsdam, NY 13699, USA}} 
\\
\\
{\authornext{Jie Sun}}\\
{\aff{Department of Mathematics and Computer Science, Clarkson University, Potsdam, NY 13699, USA}} \\
{\aff{Institute for a Sustainable Environment, Clarkson University, Potsdam, NY 13699, USA}}
\\
\\
{\authornext{Adilson E. Motter}}\\
{\aff{Department of Physics and Astronomy, Northwestern University, Evanston, IL 60208, USA}} 
\\
{\aff{Northwestern Institute on Complex Systems, Northwestern University, Evanston, IL 60208, USA}}}

\date{}% No date.

\abstract{Real-world systems are often complex, dynamic, and nonlinear. Understanding the dynamics of a system from its observed time series is key to the prediction and control of the system's behavior. While most existing techniques tacitly assume some form of stationarity or continuity, abrupt changes, which are often due to external disturbances or sudden changes in the intrinsic dynamics, are common in time series. Structural breaks, which are time points at which the statistical patterns of a time series change, pose considerable challenges to data analysis. Without identification of such break points, the same dynamic rule would be applied to the whole period of observation, whereas false identification of structural breaks may lead to overfitting. In this paper, we cast the problem of decomposing a time series into its {\it trend} and {\it seasonal} components as an optimization problem. This problem is ill-posed due to the arbitrariness in the number of parameters. To overcome this difficulty, we propose the addition of a penalty function (i.e., a {\it regularization} term) that accounts for the number of parameters. Our approach simultaneously identifies seasonality and trend without the need of iterations, and allows the reliable detection of structural breaks. The method is applied to recorded data on fish populations and sea surface temperature, where it detects structural breaks that would have been neglected otherwise. This suggests that our method can lead to a general approach for the monitoring, prediction, and prevention of structural changes in real systems.}

\maketitle

\section{Introduction}
\label{sec:intro}
Systems in the real world are often complex not only with respect to the underlying interaction networks but also with respect to their dynamics.
Examples include the temporal variations in economic growth~\shortcite{Cerra2008AER}, fluctuations of metabolic rates~\shortcite{Labra2007PNAS}, oscillations in power-grid generators~\shortcite{Filatrella2008EPJB}, and dynamics of animal species~\shortcite{Bjornstad2001Science}.
Understanding the dynamics from limited information (often in the form of a time series) is key for the prediction and control of system-level behavior.

\begin{figure}[htbc]
\centerline{
\includegraphics[width=3.5in]{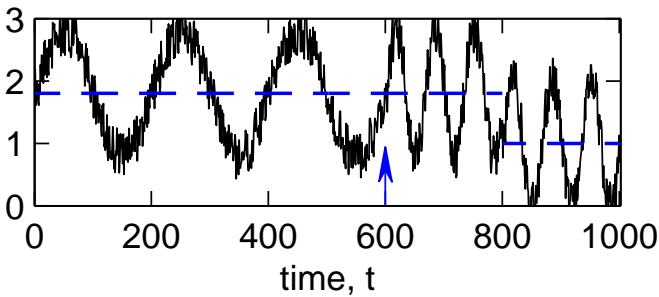}
} \caption{Example of a noisy time series that exhibits structural changes in both the trend component (dashed lines) and seasonal component (arrow).} 
\label{fig:1}
\end{figure}

While time series analysis can benefit from modern statistics, dynamical systems, and network theory, most existing techniques tacitly assume some form of stationarity or continuity~\shortcite{Brockwell2005}. However, abrupt changes, which we refer to as {\it structural breaks}, are quite common in time series (Figure~\ref{fig:1}). These are often due to external disturbances or sudden changes in the intrinsic dynamics and pose considerable
challenges to data analysis and interpretation.

Structural breaks are points in time at which the statistical patterns of a time series change~\shortcite{Andrews1993,Bai2003}. Without identification of such break points, the same dynamic rule would be applied to the whole period of observation, resulting in biases in the estimation of the system dynamics. On the other hand, false identification of structural breaks may split the time period into unnecessarily small subintervals, which affects statistical significance, introduces unnecessary parameters,  and may lead to overfitting.

In this paper, we cast the problem of decomposing a time series into its {\it trend} and {\it seasonal} components (traditionally achieved by an iterative scheme) as an optimization problem, whose objective function is the norm of the residuals. We show that this problem is ill-posed due to the arbitrariness in the number of parameters. To overcome this difficulty, we propose the addition of a penalty function (i.e., a {\it regularization} term) that accounts for the number of parameters---a strategy often used in dealing with ill-conditioned linear systems~\shortcite{Neumaier1998SIAM}. This modest change leads to successful identification of seasonality and trend without the need of iterations.
Furthermore, we show that our formulation allows the simultaneous decomposition of a time series into a trend component, a seasonal component, and a noise component, as well as structural changes in these components. Our approach is therefore a generalization of the classical method of Bai and Perron~\shortcite{Bai1998,Bai2003}, which only deals with time series without a seasonal component, and is an alternative to the recently proposed methods for the identification of structural breaks in the trend of a seasonal time series~\shortcite{Haywood2008,Verbesselt2010}.

We validate our method using synthetic data and apply it to time series describing fish populations and sea surface temperature. We found structural breaks that would have been neglected using previous methods. This indicates that our method is promising in the development of improved approaches for the monitoring, prediction, and prevention of structural changes in real systems.

\section{General Structural Break Model}
The temporal dynamics of a system is often represented by a time series $\{Y_{t}\}_{t=1}^{T}$, where $Y_t\in\mathbb{R}$ is the state of the observed variable at time $t$. The classical seasonal-trend decomposition of a time series $Y_t$ can be expressed as~\shortcite{Cleveland1990STL,Brockwell2005}
\begin{equation}\label{eq:general}
	Y_t = \mathcal{T}_t + \mathcal{S}_t + \mathcal{E}_t,
\end{equation}
where $\mathcal{T}_t$ is the trend component (usually modeled as a function of $t$), $\mathcal{S}_t$ is the seasonal component ($\mathcal{S}_{t+d}=\mathcal{S}_t$ where $d>0$ is the period of the component), and $\mathcal{E}_t$ is random noise (which is assumed to have zero mean).

Although Eq.~\eqref{eq:general} is generally valid, finding the decomposition itself turns out to be a nontrivial task. In particular, in the presence of structural breaks the decomposition will depend on the location of the break points, which are themselves dependent on the decomposition. This intrinsic coupling renders classical decomposition techniques~\shortcite{Cleveland1990STL,Brockwell2005} inappropriate in general.

To incorporate the presence of possible structural breaks into the trend component, we divide the time interval $[0,T]$ into $m$ subintervals according to the partition
\begin{equation}\label{eq:tbreak}
	0=t^*_0<t^*_1<\dots<t^*_m=T,
\end{equation}
and assume that the trend (as a function of time) remains the same within each subinterval $t^*_i+1\leq{t}\leq{t^*_{i+1}}$ ($i=0,1,\dots,m-1$).
Similarly, for the seasonal component, we partition the time period into $n$ subintervals according to
\begin{equation}\label{eq:sbreak}
	0=t^+_0<t^+_1<\dots<t^+_n=T,
\end{equation}
and assume that the seasonal component is unchanged within each subinterval $t^+_i+1\leq{t}\leq{t^+_{i+1}}$ ($i=0,1,\dots,n-1$).

Here we model the trend in the $i$-th trend subinterval as a linear function of time,
\begin{equation}\label{eq:trend}
	\mathcal{T}_t = a_{i}t + b_{i},~~t^*_i+1\leq t\leq t^*_{i+1}.
\end{equation}
On the other hand, the seasonal component in the $i$-th seasonal subinterval is represented by a set of $d_i$ numbers $\{s^{(i)}_1,s^{(i)}_2,\dots,s^{(i)}_{d_i}\}$, as
\begin{equation}\label{eq:seasonal}
	\mathcal{S}_t= s^{(i)}_k,~~t^+_i+1\leq t\leq{t^+_{i+1}},
\end{equation}
with $k=t-d_i\lfloor{t/d_i}\rfloor$,
where $d_i$ is the period of seasonality in the $i$-th subinterval and $\lfloor{x}\rfloor$ is defined as the largest integer that is smaller than or equal to $x$.

\section{Seasonal-Trend Decomposition}
We first consider the problem of seasonal-trend decomposition of a time series $\{Y_t\}_{t=1}^{T}$ in the absence of structural breaks. In particular, the goal is to decompose $\{Y_t\}_{t=1}^{T}$ as in Eq.~\eqref{eq:general} and at the same time estimate the parameters $a$ and $b$ in Eq.~\eqref{eq:trend} for the trend component and the parameters $\{s_1,s_2,\dots,s_d\}$ in Eq.~\eqref{eq:seasonal} for the seasonal component.

\subsection{Traditional Approach}
\label{sec:stdclassical}
The traditional approach for seasonal-trend decomposition often involves three steps~\shortcite{Brockwell2005}. 

The first step is to estimate the trend $\mathcal{T}_t$ by applying a moving average filter of size $q$ that attempts to eliminate the seasonal component and meanwhile reduces noise. The resulting series is computed as
\begin{equation}\label{eq:ma}
	\hat{\mathcal{T}}_t = \frac{1}{q}\sum_{j=t-\lfloor{q/2}\rfloor}^{t+\lfloor{q/2}\rfloor}w_jY_{j},
\end{equation}
where the coefficients are $w=[1,1,\dots,1,1]$ if $q$ is odd and $w=[0.5,1,\dots,1,0.5]$ if $q$ is even. 
If the period $d$ is given, a natural choice would be $q=d$.

The second step is to estimate the seasonal component. For each $k=1,2,\dots,q$, we compute $s_k$ as
\begin{equation}
	s_k = \langle Y_{k+jq} \rangle,
\end{equation}
where $\langle\cdot\rangle$ indicates average, and $j$ takes all integer values for which $1\leq{k+jq}\leq{T}$. Once the values of $s_k$ are calculated, we obtain the seasonal component $\mathcal{S}_t$ according to Eq.~\eqref{eq:seasonal} for the entire time period.

The third step is to re-estimate the trend component, often by fitting a least squares polynomial to the series $\{Y_t-\mathcal{S}_t\}$. The resulting polynomial is taken as the trend component $\mathcal{T}_t$, and its least squares polynomial fit gives the model parameters used in Eq.~\eqref{eq:trend}.

\begin{figure}[htbc]
\centerline{
\includegraphics[width=3.5in]{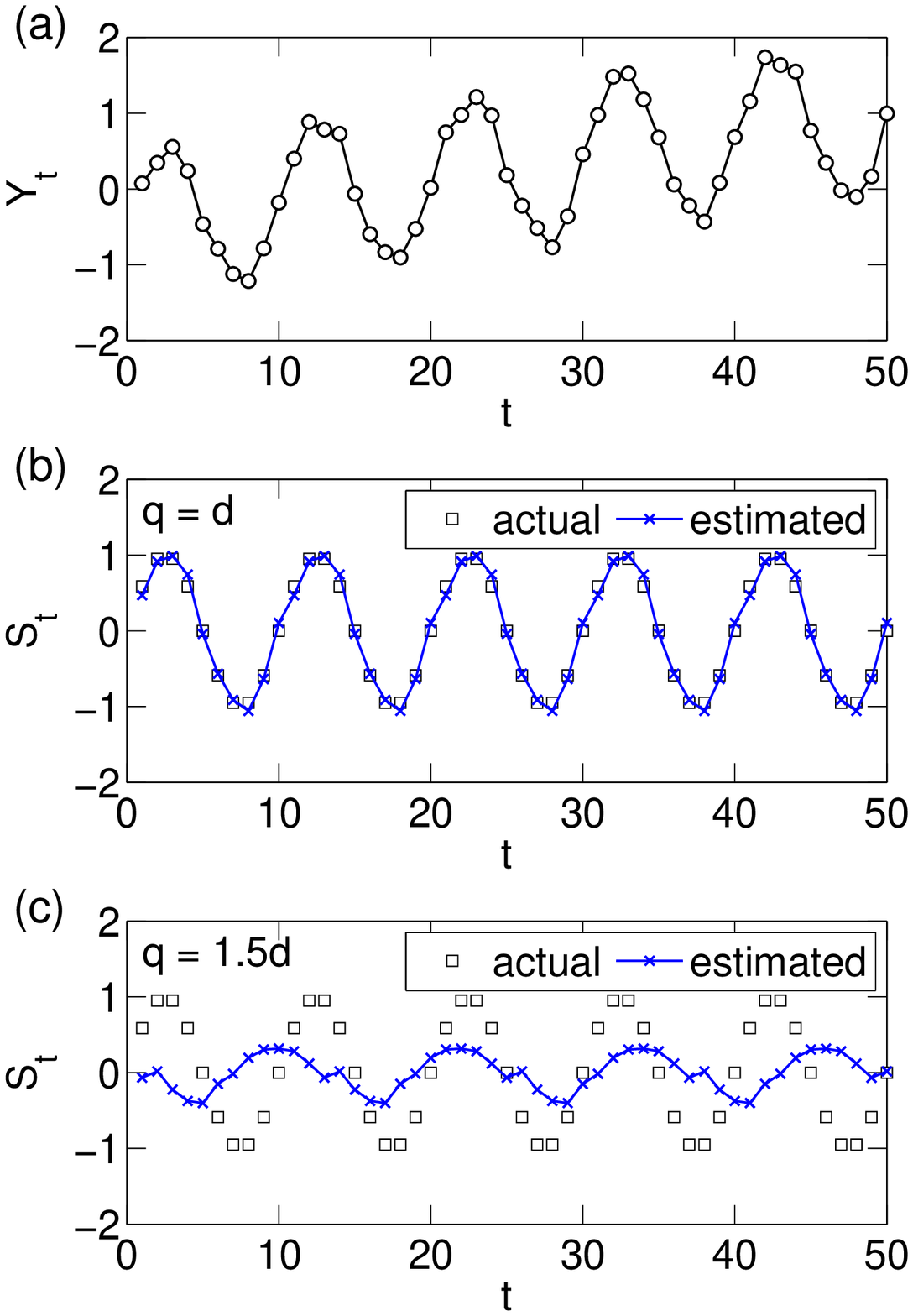}
} \caption{Classical seasonal-trend decomposition.
(a) Synthetic time series generated by the equation $Y_t=at+b+\sin(2\pi{t}/d)+\varepsilon_t$, where $a=0.03$, $b=-0.5$, $d=10$, and each $\varepsilon_t$ is generated independently from a Gaussian distribution with mean $\mu=0$ and standard deviation $\sigma=0.1$.
(b) Actual seasonal component $\mathcal{S}_t=\sin(2\pi{t}/d)$ and its estimated counterpart by the traditional approach described in Section~\ref{sec:stdclassical}, with moving average filter size $q=d$ in Eq.~\eqref{eq:ma}.
(c) Same as in panel (b), but now for $q=1.5d$.} 
\label{fig:2}
\end{figure}

In Figure~\ref{fig:2} we show that the above approach is effective when the size of the moving average filter is chosen to be $q=d$, and may generate misleading results when a different value of $q$ is used. Since $d$ is unknown in general, this approach often requires trial and error before an appropriate decomposition can be achieved.
We note that there are other methods to perform seasonal-trend decompositions, including the STL procedure proposed by~\shortciteN{Cleveland1990STL}. Most of the more sophisticated methods require iterations of multiple steps, and yet do not guarantee a reliable decomposition into seasonal and trend components.

\subsection{Regularized Optimization Approach}\label{sec:3.2}
We propose an optimization-based approach for the seasonal-trend decomposition. The idea is that, for a given trend model (e.g., the linear model in Eq.~\eqref{eq:trend}) and estimated seasonal period $p$, we have the following matrix equation for $Y={[Y_1,Y_2,\dots,Y_T]}^\intercal:$
\begin{eqnarray}\label{eq:STDgeneral}
	Y &=& Q\delta + \mathcal{E} \nonumber\\
	 &=& \begin{bmatrix}
  1 & 1 & 1 & \ldots & 0\\
  \vdots & \vdots & \vdots & \ddots & \vdots\\
  p & 1 & 0 &\ldots & 1 \\
  p+1 & 1 & 1 & \ldots & 0\\
  \vdots & \vdots & \vdots & \ddots & \vdots\\
  2p & 1 & 0 &\ldots & 1 \\
  \vdots & \vdots & \vdots & \ddots & \vdots
 \end{bmatrix}
 \begin{bmatrix}
	a \\
	b \\
	s_1 \\
  	s_2 \\
	\vdots \\
	s_{p}
 	\end{bmatrix}
	+\mathcal{E},
\end{eqnarray}
where $Q$ is a $T\times{(p+2)}$ constant matrix, $\delta$ is a $(p+2)\times{1}$ vector of parameters, and $\mathcal{E}$ is a $T\times{1}$ residual vector.
The problem of decomposing the time series into its trend and seasonal components then becomes the problem of estimating the parameter vector $\delta$ in Eq.~\eqref{eq:STDgeneral}. A common criterion is to minimize the sum of the squares of the residuals, $\|\mathcal{E}\|^2$. For fixed $p$ and Euclidean norm, $\|\cdot\|_2$, this criterion leads to the least squares estimate:
\begin{equation}
	\delta = Q^{+}Y,
\end{equation}
where $Q^{+}$ is the {\it pseudo-inverse} of the matrix $Q$~\shortcite{Horn1985}.

Note that, if we are allowed to freely choose $p$, the choice $p=T$ and $\delta={[0,c,{Y}^\intercal-c]}^\intercal$ would yield $\mathcal{E}=0$ for any number $c$, which corresponds to the minimal possible sum of squared residuals. In other words, the problem of minimizing $\|\mathcal{E}\|$ has an infinite number of solutions and is therefore {\it ill-posed} as is. 

To overcome this difficulty, we propose the addition of a penalty function to the objective function. For the given $Y$, we solve the regularized optimization problem
\begin{equation}\label{eq:STDopt}
%\underset{\delta\in\mathbb{R}^{p+2},~0\leq{p}\leq{T}}{\text{minimize}}
\underset{\delta\in\mathbb{R}^{p+2},~0\leq{p}\leq{T}}{\min}
J(\delta) = \|Y-Q\delta\|_2 + \lambda\|\delta\|_0,
\end{equation}
where $Q$ is the matrix defined in Eq.~\eqref{eq:STDgeneral}, $\lambda>0$ is a (predefined) regularization parameter, and $\lambda\|\delta\|_0$ is the penalty term where $\|\delta\|_0$ accounts for the number of parameters in the model.

For a given regularization parameter $\lambda$, the optimization problem~\eqref{eq:STDopt} can be solved in two steps:
\begin{equation}
\begin{cases}
	\mbox{Step 1:} & \mbox{For each $p$, find the solution $\delta^{(p)}$ }\\
	& \mbox{that minimizes $J(\delta)$ for $\lambda=0$.}\\
	\mbox{Step 2:} & \mbox{Choose $\delta=\argmin_{0\leq{p}\leq{T}}~J(\delta^{(p)})$}\\
	& \mbox{for the given $\lambda$.}
\end{cases}
\end{equation}
%$\delta=\underset{0\leq{p}\leq{T}}{\argmin}~J(\delta^{(p)})$
In practice, the regularization parameter $\lambda$ is often chosen to be a positive number that is small relative to the sum of squared residuals term in Eq.~\eqref{eq:STDopt}.

\begin{figure}[htbc]
\centerline{
\includegraphics[width=3.5in]{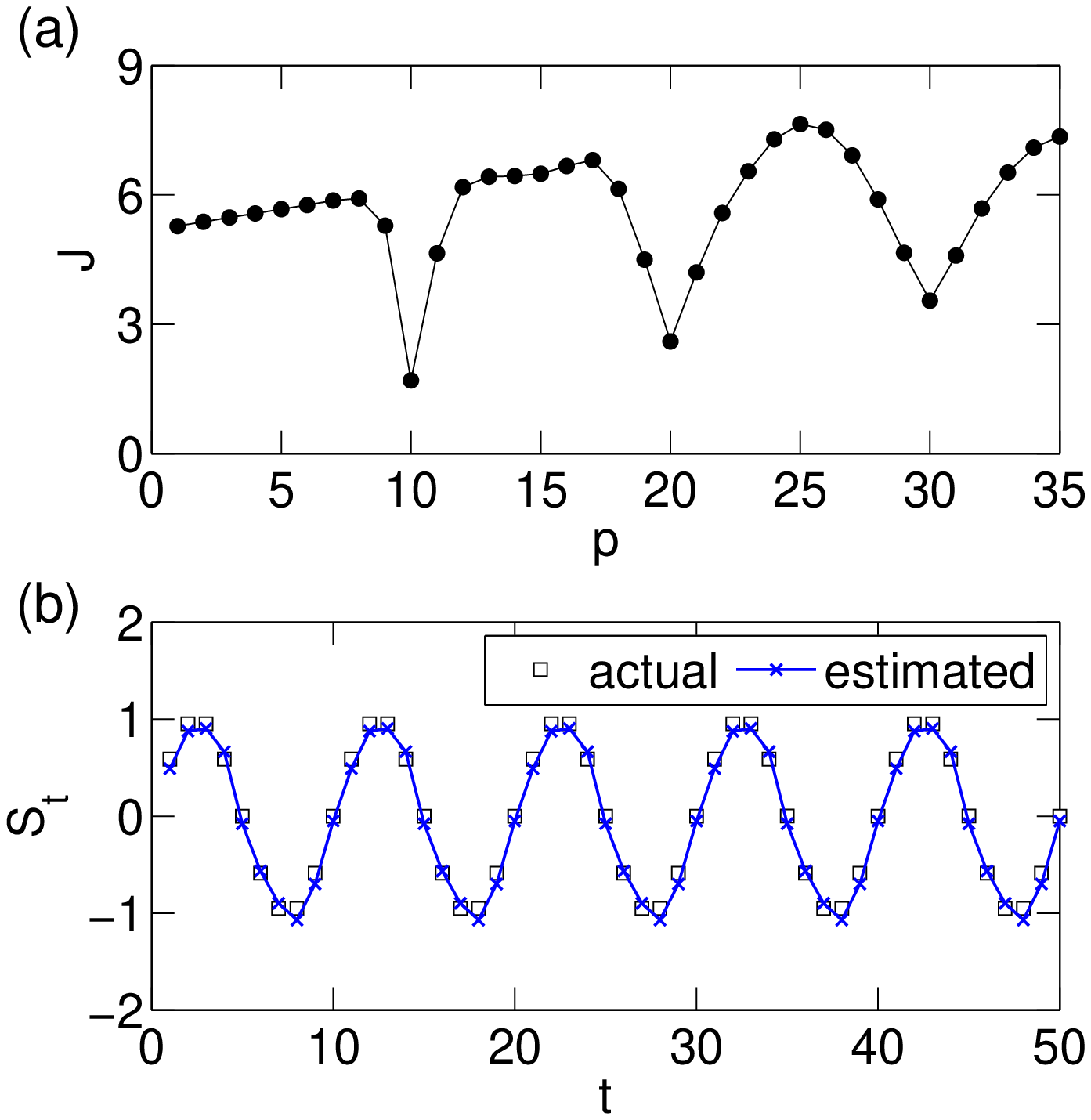}
}\caption{Seasonal-trend decomposition by regularized optimization. The time series is the same shown in Fig.~\ref{fig:2}(a).
(a) Dependence of the objective function $J(\delta^{(p)})$ on $p$, where $J$ is defined in Eq.~\eqref{eq:STDopt} for $\lambda=0.1$ and $\delta^{(p)}$ denotes the optimal solution for the given $p$. Function $J$ attains its minimum when $p=d=10$, which is the true period of the seasonal component in this example.
(b) Actual seasonal component $\mathcal{S}_t=\sin(2\pi{t}/d)$ and its estimated counterpart by the regularized optimization approach summarized in Eqs.~(10-11).} 
\label{fig:3}
\end{figure}

We apply this regularized optimization approach to the synthetic time series used in Fig.~\ref{fig:2}. Figure~\ref{fig:3}(a) shows that the regularized optimization successfully detects the true period of the seasonal component when $\lambda=0.1$. In fact, the minimal value of $J$ is achieved at $p=d=10$ for any $\lambda\in(\lambda_{\min},\lambda_{\max})$, where, in this example, $\lambda_{\min}\approx0.01$ and $\lambda_{\max}\approx0.5$. This suggests robustness of the regularized optimization approach with respect to the regularization parameter $\lambda$. Figure~\ref{fig:3}(b) shows the excellent agreement between the estimated seasonal component and the actual data (the fit to the trend, which is not plotted, is also excellent for this example).

\section{Structural Breaks in Trend}
We now turn to a slightly different problem, where the goal is to identify structural breaks in the trend component in the absence of seasonality. In particular, we assume that $\mathcal{S}_t=0$ in Eq.~\eqref{eq:general}, but $m>1$ in Eq.~\eqref{eq:tbreak}. In this case, the model for $Y$ becomes
\begin{eqnarray}\label{eq:PureBreak}
	Y&\hspace{-0.1in}=\hspace{-0.1in}& Q\delta + \mathcal{E} \\
	&\hspace{-0.1in}=\hspace{-0.1in}& \begin{bmatrix}
     1 & t^*_0+1 & &\\
     \vdots & \vdots & &\\
     1 & t^*_1 & & & \\
     & \ddots & \ddots&  & &\\
     & & & 1 & t^*_{m-1}+1 \\
     & & & \vdots & \vdots\\
     & & & 1 & t^*_{m}
 \end{bmatrix}
 \begin{bmatrix}
	a_0 \\
	b_0 \\
	a_1 \\
	b_1 \\
	\vdots\\
	a_{m-1}\\
	b_{m-1}
 	\end{bmatrix}
	+\mathcal{E},\nonumber
\end{eqnarray}
where $Q$ is $T\times{2m}$ and $\delta$ is $2m\times{1}$.
For a given time series $\{Y\}_{t=1}^{T}$, the ultimate goal within the modeling framework of Eq.~\eqref{eq:PureBreak} is to estimate the number of segments $m$, the break points $\{t^*_k\}_{k=1}^{m-1}$, and the corresponding parameters $(a_k,b_k)$ within each segment.

\subsection{Dynamic Programming Approach}\label{sec:dpa}
Model~\eqref{eq:PureBreak} has the same form as the pure structural change model~\shortcite{Bai1998}. 
For a fixed number of segments $m$, an efficient way to find the break points and the corresponding parameters that minimize the sum of residual squares is to use dynamic programming~\shortcite{Bai2003}. Let $\mbox{SSR}(i,j)$ be the sum of squared residuals obtained by applying least-squares to a segment that starts at $t=i$ and ends at $t=j$, i.e., 
\begin{equation}
	\mbox{SSR}(i,j) = \min_{a,b}\sum_{t=i}^{j}(Y_t-a-bt)^2.
\end{equation}
Furthermore, let $\mbox{SSR}(\{\tau;k\})$ be the minimum sum of squared residuals for the first $\tau$ values of $Y$ using $k$ breaks. The desired solution $\mbox{SSR}(\{T;m\})$ satisfies the following recursive equation
\begin{eqnarray}
	&&\mbox{SSR}(\{T;m\}) = \nonumber\\
	&&~~\min_{1\leq{j}\leq{T}}[\mbox{SSR}(\{j;m-1\}) + \mbox{SSR}(j+1,T)].
\end{eqnarray}
Note that in the calculation of $\mbox{SSR}(i,j)$, one can use the recursive relation that relates $\mbox{SSR}(i,j)$ to $\mbox{SSR}(i,j-1)$ for computational efficiency~\shortcite{Brown1975}. In a time series of length $T$, the dynamic programming algorithm involves order $T^2$ operations. Although not to be discussed in this paper, we point out that there has been recent work addressing in detail computational aspects of the dynamic programming algorithm and how its efficiency can be improved~\shortcite{Rigaill2010}.

\subsection{Choosing the Number of Breaks via Regularized Optimization}\label{sec:dpa2}
The dynamic programming approach requires the number of breaks $m$, which in general cannot be determined {\it a priori}. Instead of relying on statistical tests~\shortcite{Zeileis2003,Zeileis2005}, which often assume a specific form for the distribution of the residual series, here again we treat the problem (of determining the number and location of structural breaks) as a regularized optimization problem:
\begin{equation}\label{eq:optbreak}
\underset{\{t^*_k\}_{k=1}^{m},1\leq{m}\leq{T/2}}{\text{minimize}}
\Big(\big[\mbox{SSR}(\{T;m\})\big]^{1/2} + 2m\lambda\Big),
\end{equation}
where the $\mbox{SSR}$ term accounts for the sum of squared residuals, $\lambda>0$ is a regularization parameter, and $2m$ accounts for the total number of parameters when the time series is partitioned into $m$ segments.

\subsection{Example: Fish Populations in Green Bay}
We apply the regularized optimization approach to the population abundances of $43$ fish species in the Green Bay, Wisconsin. The time series for each species contains $30$ data points, corresponding to the annual abundance of that species from $1980$ through $2009$.

\begin{figure}[htbc]
\centerline{
\includegraphics[width=3.5in]{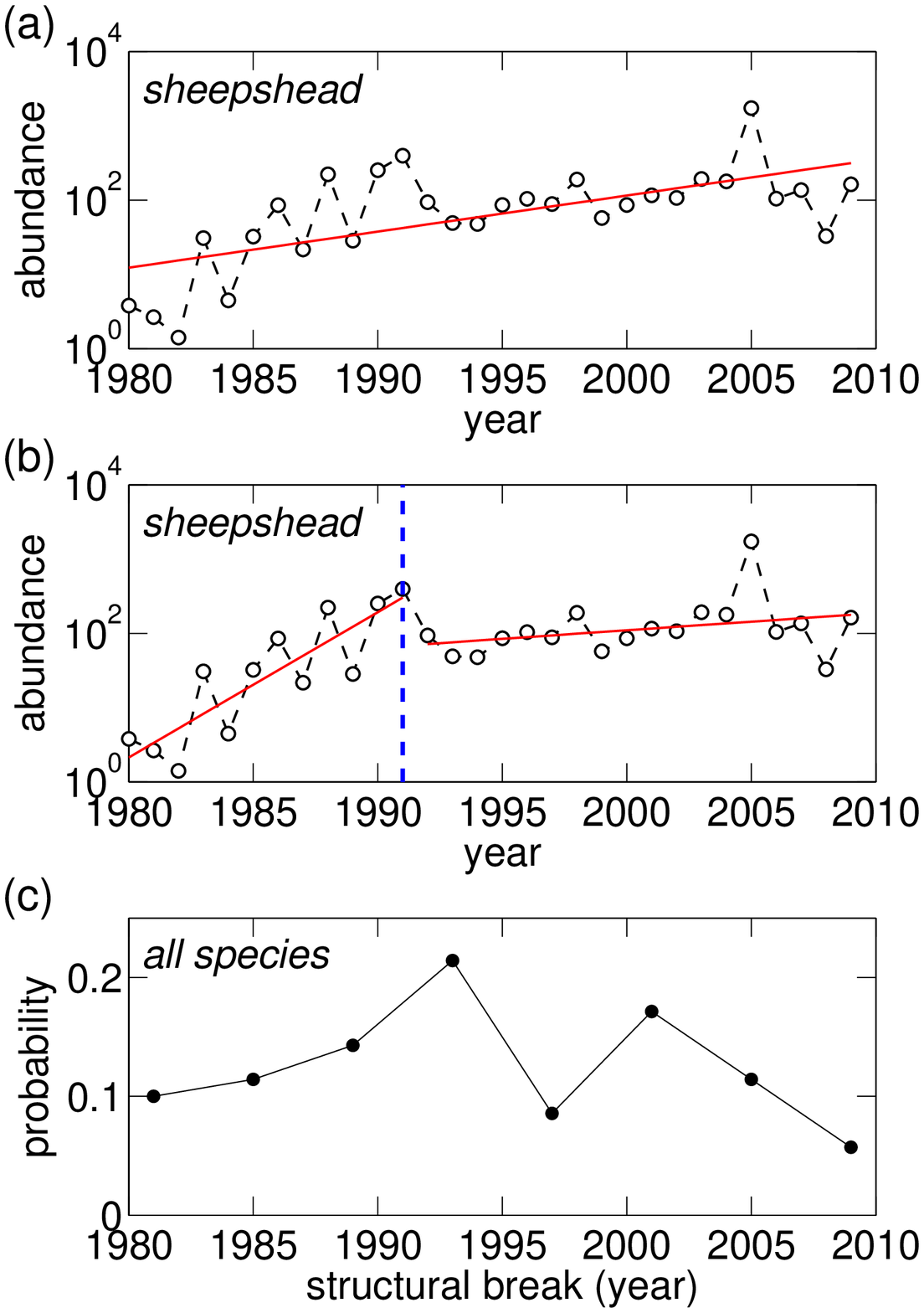}
} \caption{Structural breaks in the Green Bay fish populations.
(a) Annual population abundance of sheepshead, a fish species from Green Bay. The solid line shows the least squares fit of the time series.
(b) Same time series as in (a), partitioned into two segments using the dynamic programming algorithm described in Section~\ref{sec:dpa}. The vertical dashed line marks the break point and the two solid lines represent the least squares fit for the two time series segments, respectively.
(c) Probability distribution of break points for all $43$ species identified using our method. Note the peaks around $1993$ and $2001$ of this distribution.}\vspace{-0.3in}
\label{fig:4}
\end{figure}

Figure~\ref{fig:4}(a) shows the population abundance of the fish sheepshead and its best linear trend. Using $\lambda=0.15$ in Eq.~\eqref{eq:optbreak}, we find that the optimal solution requires $m=2$ segments, with the structural break occurring at $t^*_1=12$ (year $1991$), as shown in Fig.~\ref{fig:4}(b). Furthermore, we apply Eq.~\eqref{eq:optbreak} to all $43$ species with $\lambda=0.15$ and obtain a distribution of the break times for all species as shown in Fig.~\ref{fig:4}(c). It is interesting to observe that the distribution curve has two peaks: one in the early 1990's which is the time the invasive species round goby was first discovered in the system, and another peak in the early 2000's~\shortcite{Lederer2006,Lederer2008}.\vspace{-0.2in}

\section{Structural Breaks in the Trend of a Seasonal Time Series}
We now present our approach to the problem of detecting structural breaks in a time series that contains both trend and seasonal components. Such a time series can be decomposed using the general form in Eq.~\eqref{eq:general} with the possible presence of structural breaks defined by Eqs.~(\ref{eq:tbreak}-\ref{eq:seasonal}).

\subsection{Regularized Optimization Formulation}
Using the notation introduced in Eqs.~(\ref{eq:general}-\ref{eq:seasonal}) for the general seasonal time series with structural breaks, the time series model for $Y$ can be expressed as
\begin{equation}\label{eq:16}
	Y = Q_{\mathcal{T}}\delta_{\mathcal{T}} + Q_{\mathcal{S}}\delta_{\mathcal{S}} + \mathcal{E}.
\end{equation}
Here, the term
\begin{eqnarray}\label{eq:17}
	Q_{\mathcal{T}}\delta_{\mathcal{T}}=\hspace{-0.05in}\begin{bmatrix}
     1 & t^*_0+1 & &\\
     \vdots & \vdots & &\\
     1 & t^*_1 & & & \\
     & \ddots & \ddots&  & &\\
     & & & 1 & t^*_{m-1}+1 \\
     & & & \vdots & \vdots\\
     & & & 1 & t^*_{m}
 \end{bmatrix}
 \begin{bmatrix}
	a_0 \\
	b_0 \\
	a_1 \\
	b_1 \\
	\vdots\\
	a_{m-1}\\
	b_{m-1}
 	\end{bmatrix}
\end{eqnarray}
denotes a linear model for the trend component with $m-1$ structural breaks ($m$ segments). The term
\begin{eqnarray}\label{eq:18}
  Q_{\mathcal{S}}\delta_{\mathcal{S}}    = \begin{bmatrix}
  Q_{\mathcal{S}}^{(1)} & & & \\
  & Q_{\mathcal{S}}^{(2)} & & \\
  & & \ddots & \\
  & & & Q_{\mathcal{S}}^{(n)}
 \end{bmatrix}_{T\times\sum_{d_i}}\hspace{-0.05in}
 \begin{bmatrix}
	s^{(1)} \\
  	s^{(2)} \\
	\vdots \\
	s^{(n)}
 	\end{bmatrix}_{\sum_{d_i}\times{1}}\hspace{-0.3in}
\end{eqnarray}
models the seasonal component with $n-1$ breaks ($n$ segments), where
\begin{eqnarray}\label{eq:19}
  Q_{\mathcal{S}}^{(i)}s^{(i)}    = \begin{bmatrix}
  1 & \ldots & 0\\
  \vdots & \ddots & \vdots\\
  0 &\ldots & 1 \\
  1 & \ldots & 0\\
  \vdots & \ddots & \vdots\\
  0 &\ldots & 1 \\
  \vdots & \ddots & \vdots
 \end{bmatrix}_{(t^{+}_i-t^{+}_{i-1})\times{d_i}}\hspace{-0.5in}
 \begin{bmatrix}
	s^{(i)}_1 \\
  	s^{(i)}_2 \\
	\vdots \\
	s^{(i)}_{d_i}
 	\end{bmatrix}_{\sum_{d_i}\times{1}}\hspace{-0.3in}
\end{eqnarray}
is the seasonal model for the $i$th segment ($i=1,2,\dots,n$).

For the given $Y$, and regularization parameter $\lambda>0$, our approach is based on solving the following regularized optimization problem to find the break points as well as the corresponding parameters $\delta_{\mathcal{T}}$ and $\delta_{\mathcal{S}}$ defined in Eqs.~\eqref{eq:16}$\sim$\eqref{eq:19}:
\begin{equation}\label{eq:generalSTbreak}
%\underset{\delta\in\mathbb{R}^{p+2},~0\leq{p}\leq{T}}{\text{minimize}}
\underset{\delta\in\mathbb{R}^{p+2},~0\leq{p}\leq{T}}{\min}
	J(\delta) = \|Y-Q\|_2 + \lambda\|\delta\|_0,
\end{equation}
where $Q$ and $\delta$ are defined as
\begin{equation}
	Q = [Q_{\mathcal{T}},Q_{\mathcal{S}}],~~~~\delta = {[{\delta_{\mathcal{T}}}^\intercal,{\delta_{\mathcal{S}}}^\intercal]}^\intercal.
\end{equation}

Although the global optimum of problem~\eqref{eq:generalSTbreak} can be found by enumeration of all possible break points by solving a least squares problem for each set of break points, such a brute-force approach is not practical due to its high computational cost. Here we focus on an alternative method, which is based on iterative optimization of the trend and seasonal parameters, respectively. In particular, we consider the scenario where the seasonal period $d$ does not change, but allows for the presence of seasonality, trend, and structural breaks in the trend component. Under such conditions, the following procedure is adopted for the detection of structural breaks and estimation of parameters (including those for the seasonal component):
\begin{equation}
\begin{cases}
	\mbox{Step 0:} & \mbox{Initial assignment of $\mathcal{T}=0$.}\\
	\mbox{Step 1:} & \mbox{Seasonal-trend decomposition of $Y-\mathcal{T}$}\\
	& \mbox{via the methods from Sec.~\ref{sec:3.2} to obtain $\mathcal{S}$.}\\
	\mbox{Step 2:} & \mbox{Estimation of the structural breaks and}\\
	& \mbox{parameters of $Y-\mathcal{S}$ using the methods}\\
	& \mbox{described in Secs.~\ref{sec:dpa}--\ref{sec:dpa2}.}\\
	& \mbox{Update the estimated trend as $\mathcal{T}=Q_{\mathcal{T}}\delta_{\mathcal{T}}$.}
\end{cases}
\end{equation}
Steps $1$ and $2$ are repeated until the estimation of the breaks points and trend converges.

\subsection{Example: Arctic Sea Surface Temperature}
We apply our method to the time series of the sea surface temperature (SST) from the Arctic. The time series data is obtained from the U.S. Geological Survey website~\shortcite{USGS2013}. The original data contains moderate-resolution monthly SST from $20$ regional seas in the Arctic over $28$ years (1982--2009). 

\begin{figure}[htbc]
\centerline{
\includegraphics[width=3.5in]{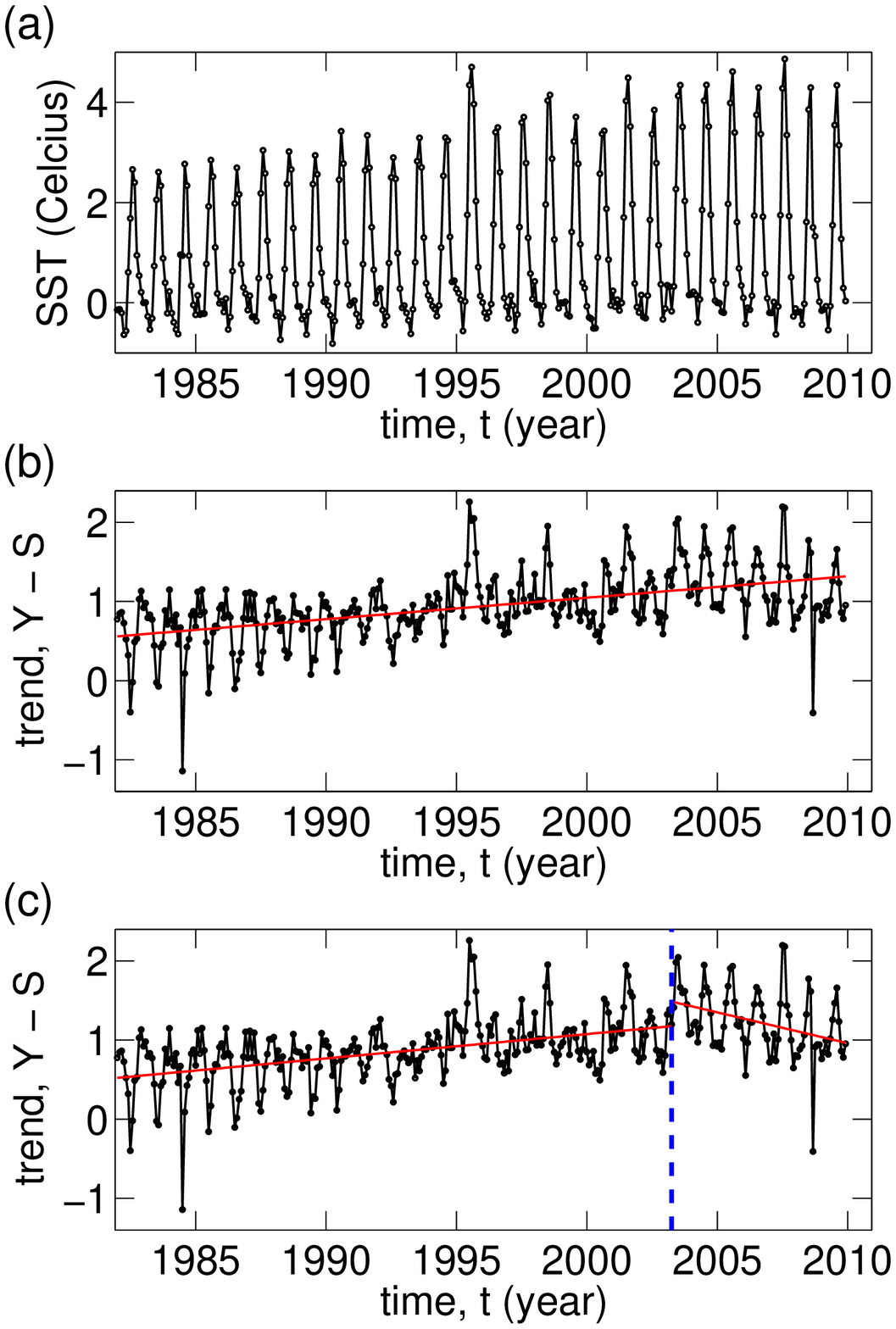}
} \caption{Structural break in the Arctic sea surface temperature time series. 
(a) Monthly SST from January $1983$ through December $2009$.
(b) Seasonal-adjusted time series, $Y-\mathcal{S}$, as well as its linear dependence of time (solid straight line), which resembles the global trend of the time series.
(c) Same as (b), except that a break point around 2003 is identified using our approach (vertical dashed line). The break point separates the seasonal-adjusted time series into two temporal regimes, one slowly increasing in time and the other slowly decreasing in time (solid straight lines on the two sides of the break point).
}
\label{fig:5}
\end{figure}

We focus on the overall monthly SST of the Arctic obtained by averaging over the SST from all $20$ regional seas. The data for several regional seas are not available for the entire time period. To address this issue, we fill in the missing data by the temporal average SST from available data for that sea. After this pre-processing step, we obtain the time series $\{Y_t\}_{t=1}^{T}$, as shown in Fig.~\ref{fig:5}(a), where $Y_t$ denotes the average SST of the Arctic during the $t$-th month ($t=1$ corresponds to January 1983 and $T=336$).

Using a regularization parameter $\lambda=0.1$, we found that the optimal seasonal period for the time series is $d=12$, with the presence of $1$ break point in the trend component, around the month of April in year 2003. Figures~\ref{fig:5}(b-c) show the seasonal-adjusted time series as well as their trend component when there is no structural break and when there is one structural break (found to yield optimal solution via our approach). Note that the current trend of the Arctic SST would have been asserted to be increasing if structural break in the trend component were ignored. However, our result suggests that, despite the overall increase for $20$ years (1982-2001), the recent trend of the SST (2002-2009) is decreasing rather than increasing.

\section{Conclusions}
In this paper, we proposed a regularized optimization framework for time series analysis, including renewed approaches for solving the classical problem of seasonal-trend decomposition and the detection of structural breaks in a time series with or without the presence of seasonal components.
Our approach was tested against synthetic data and applied to empirical time series, including fish populations from Green Bay and sea surface temperature in the Arctic. We are able to detect structural breaks that are of practical importance in the prediction of future states of these complex systems.

The key in our approach is the formulation that naturally treats parameters for the trend and seasonal components in a similar manner. The regularization term is analogous to the Akaike information criterion~\shortcite{Akaike1974} and the Schwarz-Bayesian criterion~\shortcite{Schwarz1978}, both of which are popular methods for model selection. In our formulation, these methods can be useful for the selection of regularization parameters. Other approaches, such as the detrended fluctuation analysis~\shortcite{Hu2001PRE,Chen2002PRE}, are also likely to be useful when integrated into our framework.

We used Euclidean norm to measure the model quality, due to its analytical convenience (solution to least square problems can be found easily using singular value decomposition). In view of the recent developments in parameter estimation under general $p$ norms, including $p\leq1$ for sparse data~\shortcite{Candes2006}, and $p=\infty$ when the underlying dynamics is chaotic~\shortcite{Sun2011}, it will be interesting to explore such scenarios and how they can improve the estimation of break points. Faster computational procedures, however, will be necessary since most such problems involve much higher computational cost than ordinary least square problems.

Finally, the problem of predicting future occurrences of structural breaks---or {\it tipping points}, in the language of climatology and ecology---remains a challenging problem~\shortcite{Scheffer2009}. It is our hope that the development of new and improved methods for the prediction of future trends, seasonal patterns, and breaks will benefit from from methods (such as the one proposed here) to  estimate those quantities reliably in recorded time series.

\section*{Acknowledgements}
The authors thank John Janssen for providing data on fish populations.
This research is based on work supported by NSF (Grant No.~1060382), ARO (Grant No.~61386-EG), and NOAA (Grant No.~NA09NMF4630406).

\bibliographystyle{chicaco}
\bibliography{2014_StructuralBreakOptimization.bib}

\end{document}